\documentstyle[12pt]{article}
  \advance\oddsidemargin  by -1in
  \advance\evensidemargin by -1in
  \advance\topmargin      by -0.5in
\makeindex

\def\br{{\bf r}}
\def\brp{{\bf r^{\prime}}}
\def\Pom{{\bf I\!P}}
\def\lsim{\mathrel{\rlap{\lower4pt\hbox{\hskip1pt$\sim$}}
    \raise1pt\hbox{$<$}}}         
\def\gsim{\mathrel{\rlap{\lower4pt\hbox{\hskip1pt$\sim$}}
    \raise1pt\hbox{$>$}}}         
\def\beq{\begin{equation}}
\def\eeq{\end{equation}}
\def\bea{\begin{eqnarray}}
\def\eea{\end{eqnarray}}

\begin{document}

\vspace{1.0cm}

\begin{center}

{\Large \bf
Predictions for high-energy real and virtual photon-photon scattering
from color dipole BFKL-Regge factorization
\vspace{1.0cm}\\}

{\large \bf N.N.Nikolaev$^{1,2}$, J.Speth$^{1}$ and  V.
R.Zoller$^{3}$}\\

\vspace{0.5cm} $^{1}${ \em Institut  f\"ur
Kernphysik, Forschungszentrum J\"ulich,\\ D-52425 J\"ulich, Germany\\
E-mail: kph154@ikp301.ikp.kfa-juelich.de}\\

$^{2}${ \em L.D.Landau Institute for Theoretical Physics, Chernogolovka,
Moscow Region 142 432, Russia}\\

$^{3}${\em Institute for  Theoretical and Experimental Physics,\\
Moscow 117218, Russia\\
E-mail: zoller@heron.itep.ru}\\

\vspace{0.5cm}

{\bf Abstract}

\end{center}
High-energy virtual photon-virtual photon scattering can be viewed as 
interaction of small size color dipoles from the beam and target photons, 
which makes $\gamma^{*}\gamma^{*},
\gamma^{*}\gamma$ scattering at high energies (LEP, LEP200 \& NLC) an 
indispensable probe of short distance properties of the QCD pomeron 
exchange. Based on the color dipole representation, we investigate 
consequences for the $\gamma^{*}\gamma^{*},\gamma^{*}\gamma$ scattering 
of the incorporation of 
asymptotic freedom into the BFKL equation which 
makes the QCD pomeron a series 
of isolated poles in the angular momentum plane. The emerging color 
dipole BFKL-Regge factorization allows us to relate in a model-independent 
way the contributions of each BFKL pole to $\gamma^{*}\gamma^{*},\gamma^{*}
\gamma$ scattering and DIS off protons. Numerical predictions based on 
our early works on color dipole BFKL phenomenology of DIS on protons are in
a good agreement with the  experimental data on the photon
 structure function $F_{2\gamma}$ and 
most recent data on the  $\gamma^*\gamma^*$ cross section
$\sigma^{\gamma^*\gamma^*}(Y)$ from OPAL and L3
experiments at LEP200.
 We discuss the  role of non-perturbative dynamics
 and predict pronounced effect
of the Regge-factorization breaking    due to large unfactorizable
non-perturbative  corrections to the perturbative vacuum exchange.
We comment on the  salient features of the BFKL-Regge 
expansion for $\gamma^{*}\gamma^{*},\gamma^{*}\gamma$ scattering including 
the issue of decoupling of subleading  BFKL poles and 
 the soft plus rightmost hard BFKL  pole   dominance .
\vspace{1.0cm}\\


\section{Introduction}

In this note we study  scattering of virtual and real photons
\beq
\gamma^*(q) +\gamma^*(p) \to  X\,.
\label{eq:1.1}
\eeq
in the high-energy  regime of large
Regge parameter ${1/ x}$ which  depends on virtualities of photons as  
\beq
{1 \over x}={W^2+Q^2+P^2\over Q^2+P^2+\mu^2}\gg 1.
\label{eq:1.2}
\eeq
and has  correct parton model limit if either $Q^2\ll P^2$ or $P^2\ll Q^2$.
In eq.(\ref{eq:1.2}) $W^2=(q+p)^2$ is the center-of-mass energy squared
of colliding space-like photons $\gamma^*(q)$ and $\gamma^*(p)$ with
virtualities $q^2=-Q^2$ and  $p^2=-P^2$, respectively. 

The recent strong theoretical \cite{NISIUS,BART1,BART2,BRODSOP,BRRW,BADEL,KWIE}
and experimental \cite{NISIUS,L3,OPALGG,L3SIGY,OPAL,OPALY} 
(see also a compilation
in \cite{PDG}) interest in high-energy 
$\gamma^{*}\gamma^{*},\gamma^{*}\gamma,\gamma\gamma$ scattering stems 
from the fact that virtualities of
photons give a handle on the size of color dipoles in the beam and target 
photons and, eventually, on short distance properties of the QCD pomeron 
exchange. For earlier development of the subject see the pioneering paper
\cite{BUDNEV}. 

As noticed by Fadin, Kuraev and Lipatov in 1975 \cite{FKL} and 
discussed in more detail by Lipatov in \cite{LIPAT86} the incorporation of 
asymptotic freedom into the BFKL equation \cite{BFKL} makes the QCD pomeron a series 
of isolated poles in the angular momentum plane. The contribution of each
isolated pole to the high-energy scattering amplitude satisfies the familiar
Regge factorization \cite{Gribov}. 
In \cite{JETPLett} we reformulated the consequences of
the Regge factorization in our color dipole (CD) approach to the BFKL
pomeron. In this communication we address several closely related issues
in photon-photon scattering in the Regge regime (\ref{eq:1.2}) which
can be tested at LEP200 and Next Linear Collider (NLC). 

First, following our early work \cite{JETPLett,DER,PION} we discuss how 
the color dipole (CD) BFKL-Regge factorization leads to the parameter-free
predictions for total cross sections of
 $\gamma^{*}\gamma^{*}$, $\gamma^{*}\gamma$,
$\gamma\gamma$ scattering. We find good agreement with the recent experimental 
data from the L3 and OPAL experiments at LEP \cite{L3,OPALGG,L3SIGY,OPAL,OPALY}. 

Second, we discuss the interplay of soft and hard dynamics of the vacuum exchange
and comment on the onset 
of the soft plus rightmost hard BFKL-pole dominance in $\gamma^*\gamma^*$ diffractive 
scattering. The nodal
properties of eigenfunctions of the CD BFKL equation suggest an interesting
possibility of decoupling of sub-leading BFKL singularities when the
virtuality of one or both of photons is in the broad
 vicinity of $Q^{2} \sim 20$
GeV$^{2}$. Hence,  very efficient the leading hard plus soft approximation 
(LHSA) advocated by us previously in \cite{PION}.

 Third, we discuss the impact of running CD BFKL on contentious 
issue of testing the factorization properties of photon-photon scattering 
in the $Q^{2},P^{2}$-plane which has earlier been discussed  
only in the approximation of $\alpha_S=$const to the BFKL equation
\cite{BART1,BART2,BRODSOP} (for the color dipole picture in the
 $\alpha_S=$const approximation see \cite{Mueller}). Our finding is that
the non-perturbative corrections break down the Regge factorization.
 The experimental
observation of this phenomenon would contribute to better understanding 
of the non-perturbative
dynamics of high-energy processes.

\section{Overview of color dipole BFKL-Regge factorization}

In the color dipole basis the beam-target scattering is viewed as
transition of $\gamma^*$ into quark-antiquark pair 
and interaction of the beam ($A$) and 
target ($B$) color dipoles of the flavor $A,B=u,d,s,c$. As a fundamental 
quantity we use the forward dipole scattering amplitude and/or the 
dipole-dipole cross section $\sigma(x,\bf{r},\bf{r^{\prime}})$. Once 
$\sigma(x,\br,\brp)$ is known the total cross section of $AB$ scattering 
$\sigma^{AB}(x)$ is calculated as 
\beq
\sigma^{bt}(x)
=\int dz d^{2}{\bf{r}} dz^{\prime} d^{2}{\bf{r^{\prime}}}
|\Psi_A(z,{\bf{r}})|^{2} |\Psi_{B}(z^{\prime},{\bf{r^{\prime}}})|^{2}
\sigma(x,{\bf{r},\bf{r^{\prime}}}) \,,
\label{eq:2.1}
\eeq
 where $\bf{r}$ and $\bf{r^{\prime}}$ are the 
two-dimensional vectors in the impact parameter plane.
In the color dipole factorization formula (\ref{eq:2.1}) the dipole-dipole 
cross section $\sigma(x,\bf{r},\bf{r^{\prime}})$ is beam-target symmetric
and universal for all beams and targets, the beam and target dependence is
concentrated in probabilities $|\Psi_A(z,{\bf{r}})|^{2}$ and
$|\Psi_{B}(z^{\prime},{\bf{r^{\prime}}})|^{2}$ to find a color dipole,
$\br$ and $\brp$ in the beam and target, respectively. Hereafter we focus
on cross sections averaged over polarizations of the beam and target photons,
in this case only the term $n=0$  of the Fourier series
\beq
\sigma(x,\br,\brp)=\sum_{n=0}^{\infty}\sigma_n(x,r,r^{\prime})\exp(in\varphi),
\label{eq:2.2}
\eeq
where $\varphi$ is an azimuthal angle  between $\bf r$ and $\bf{r^{\prime}}$,
contributes in (\ref{eq:2.1}).

Fadin, Kuraev and Lipatov noticed in 1975 \cite{FKL}, see also Lipatov's 
extensive discussion \cite{LIPAT86} that the incorporation of asymptotic 
freedom into the BFKL equation makes the QCD pomeron a series of isolated 
poles in the angular momentum plane. The contribution of the each pole to 
scattering amplitudes satisfies the standard Regge-factorization \cite{Gribov},
which in the CD basis implies the CD BFKL-Regge expansion for the 
vacuum exchange dipole-dipole
cross section
\beq
\sigma(x,r,r^{\prime})=\sum_{m}C_m\sigma_m(r)\sigma_m(r^{\prime})
\left({x_0\over x}\right)^{\Delta_m}\,.
\label{eq:2.3}
\eeq
Here the dipole cross section $\sigma_m(r)$ is an  eigen-function of the 
CD BFKL equation \cite{JETPLett,DER,PISMA1,NZZJETP,NZHERA}
\beq
{\partial\sigma_{m}(x,r)\over \partial \log(1/x)}=
 {\cal K}\otimes \sigma_{m}(x,r)=\Delta_{m}\sigma_{m}(x,r),
\label{eq:2.4}
\eeq
with eigen value (intercept) $\Delta_{m}$. Arguably, for transition of $\gamma^*$
into heavy flavors, $A=c,b,...$, the hardness scale is set by $Q^2+4m^2_A$,
for light flavors $Q^2+m^2_{\rho}$ is a sensible choice
 which leads to the correct value of  
 Regge parameter  in the photoproduction 
regime, $Q^2\to 0$.

 Hence,
for the light-light transition we evaluate the Regge-parameter (\ref{eq:1.2})
with $\mu^2=m^2_{\rho}$, for the light-charm contribution we take $\mu^2=4m^2_{c}$
and for the charm-charm contribution we take $\mu^2=8m^2_{c}$

For the details on CD formulation of the BFKL equation, infrared 
regularization by finite propagation radius $R_{c}$ for 
perturbative gluons and freezing of strong coupling at large distances,
the choice of the physically motivated boundary condition for the
hard BFKL evolution and description of eigenfunctions we refer to our 
early works \cite{JETPLett,DER,NZHERA},
the successful application of CD BFKL-Regge expansion to the proton and 
pion structure functions (SF) and 
evaluation of hard-pomeron contribution to the 
rise of hadronic and real photo-absorption cross sections is found in 
\cite{JETPLett,DER,PION,NZHERA}. We only recapitulate the
salient features of the formalism essential for the present discussion.

There is a useful analogy between the intercept $\Delta=\alpha(0)-1$ and
binding energy for the bound state problem for the Schr\"odinger equation. 
The eigenfunction $\sigma_0(r)$ for the rightmost hard BFKL pole (ground
state) corresponding to the largest intercept $\Delta_0\equiv\Delta_{\Pom}$
is node free. The  eigenfunctions $\sigma_m(r)$ for excited states with 
$m$ radial nodes have intercept $\Delta_{m} < \Delta_{\Pom}$. Our choice
of $R_c=0.27$fm yields for the rightmost hard BFKL pole the intercept  
$\Delta_{\Pom}=0.4$\,, for sub-leading hard poles $ \Delta_m \approx 
{\Delta_0/(m+1)}$. The  node of $\sigma_{1}(r)$ is located at
$r=r_1\simeq 0.05-0.06\,{\rm fm}$, for larger $m$ the rightmost nodes
move to a somewhat  larger $r$ and accumulate at $r\sim 0.1\, {\rm fm}$,
for the more detailed description of the nodal structure of $\sigma_m(r)$ see
\cite{JETPLett,DER}. Here we only emphasize that  for solutions with
$m\geq 3$ the third and higher nodes are located at a very small $r$  
way beyond the resolution scale $1/\sqrt{Q^2}$ of foreseeable deep inelastic
scattering (DIS) 
experiments. Notice that the Regge cut in the complex angular 
momentum plane found in the much discussed approximation 
$\alpha_{S}=$const resembles an infinite, and continuous, sequence of poles.
In the counterpart of our CD BFKL-Regge expansion (5) for approximation 
$\alpha_{S}=$const the intercept ${\Delta_m}$ would  be a continuous
 parameter in 
contrast to a discrete spectrum for standard running $\alpha_{S}$.

Because the BFKL equation sums cross sections of
production of multigluon final states, the perturbative two-gluon Born
approximation 
is an arguably natural boundary condition. This leaves the starting point
$x_{0}$ as the only free parameter which fixes completely the result of
the hard BFKL evolution for dipole-dipole cross section. We follow the 
choice $x_0=0.03$ made in \cite{NZHERA}. The very ambitious program of 
description of $F_{2p}(x,Q^2)$ starting  from  this, perhaps  excessively 
restrictive, perturbative two-gluon boundary condition has been launched 
by us in \cite{JETPLett} and met  with  remarkable phenomenological success
\cite{DER,PION}.

 Because in the attainable region of $r$ the sub-leading 
solutions $m\geq 3$ can not be resolved and they all have similar intercepts
$\Delta_m\ll 1$, in practical evaluation of $\sigma^{AB}$
 we can truncate expansion ($\ref{eq:2.3}$) at $m=3$ lumping in the term
with $m=3$ contributions of all singularities with $m\geq 3$. Specifically,
if we endow 
\beq
\sigma_3(r)=
\sigma_{Born}(r)-\sum_{m=0}^2\sigma_m(r)
\label{eq:SIGMA3}
\eeq
 with the effective intercept $\Delta_3=0.06$ the  truncated expansion
 reproduces the numerical
solution $\sigma(x,r)$ 
of our CD BFKL equation 
in the wide range of dipole sizes $10^{-3}\lsim r \lsim 10$ fm with accuracy 
$\simeq 10\%$ even at moderately small  $x$. 
 Such a
truncation can be  justified {\sl a posteriori} if such  a  contribution 
from $m\geq 3$ turns out to be a small correction, which will indeed be 
the case at  small $x$.

Whereas scattering of small dipoles $r\lsim R_c$ is  dominated by the 
exchange of perturbative gluons,  interaction of large dipoles with the proton target
 has been modeled in Ref.\cite{NZHERA,DER,PION} by the  non-perturbative
soft pomeron with intercept $\alpha_{\rm soft}(0)-1=\Delta_{\rm soft}=0$. 
Then the extra term $\sigma_{\rm soft}(r,r^{\prime})$
must be added in the  r.h.s. of expansion (\ref{eq:2.3}).

From the early phenomenology of DIS and diffractive vector meson
 production off the proton target we only know the parameterization
 of $\sigma_{\rm soft}(r,r^{\prime})$
when one of the dipoles is definitely large, of the order of the proton size.
Evaluation of the soft contribution to $\gamma^*\gamma^*$ scattering
when 
 both dipoles are small inevitably introduces model dependence. Modelling
of soft contribution by the exchange by two nonperturbative gluons suggests
\cite{Levin}
$$
 \sigma_{\rm soft}(r,r^{\prime})\propto {r^2r^{\prime 2}\over (r^2+r^{\prime 2})}
$$
and the non-factorizable cross section of the form
$$
\sigma^{\gamma^*\gamma^*}_{\rm soft}(Q^2,P^2)\propto {1\over Q^2+P^2}\,.
$$
Similar non-perturbative  $\sigma_{\rm soft}$ is found in the soft pomeron models
 \cite{NACHT}. The explicit parameterization is found in Appendix.

Finally, at moderately small values of $x$ the above described $t$-channel 
gluon tower exchange  must be complemented by the $t$-channel            
 $q\bar{q}$ exchange often associated with DIS off vector mesons (hadronic component)
and off the perturbative (point-like) $q\bar q$-component of the
 target photon wave function.
We add corresponding corrections only to the real photon structure function 
$F_{2\gamma}(x,Q^2)$ to estimate the interplay of vacuum and non-vacuum 
exchanges in the currently accessible kinematical region of not very small $x$.
 In all other cases of interst we concentrate on the pure
  vacuum exchange at $x\lsim x_0$ where the non-vacuum corrections are negligible small. 

In our evaluation of the box diagram contribution to  $F^{pl}_{2\gamma}(x,Q^2)$
 which is \cite{WZK}
\beq
F^{pl}_{2\gamma}(x,Q^2)={3\alpha_{em}\over \pi}
\sum_{q=udsc}e^4_qx\left\{\left[x^2+(1-x)^2\right]
\log{Q^2(1-x)\over xQ_q^2}+8x(1-x)-1 \right\} 
\label{eq:PL}
\eeq
we take the $\rho$-meson 
mass  as the lower cut-off for the light-flavor-loop
integral, $Q^2_q=m^2_{\rho}$ for $q=u,d,s$, and the charm quark mass
for the $c$-loop, $Q^2_c=m^2_c$. In eq.(\ref{eq:PL}) $e_q$ is a quark charge.

To describe the hadronic component of $F_{2\gamma}$ we take the
 coherent mixture of $\rho^0$ and $\omega$
mesons \cite{SaS}. Being  supplemented with quite standard assumptions on 
the vector meson valence quark density this gives  
\beq 
F^{had}_{2\gamma}(x)=
{\alpha_{em}\over 12}\left[4(g_{\rho}+g_{\omega})^2
+(g_{\rho}-g_{\omega})^2)\right]\sqrt{x}(1-x)\,,
\label{eq:HAD}
\eeq
where the coupling constants $g_V^2=4\pi/f_V^2$, in the Fock state expansion 
$$
|\gamma\rangle^{had}=
{e\over f_{\rho}}|\rho\rangle+{e\over f_{\omega}}|\omega\rangle+...
$$
 are as follows $g_{\rho}^2= 0.5$ and $g_{\omega}^2= 0.043$ \cite{PDG}.
  We neglect the $Q^2$-evolution  which, at reasonable values of lower scale, 
is a small correction on the  interval $1.9\leq Q^2\leq 5$ GeV$^2$ where
the small-$x$ data on $F_{2\gamma}$ were taken.

Then, 
combining (\ref{eq:2.3}) and (\ref{eq:2.1}) and adding in the soft and
non-vacuum  components, we obtain our principal result for
virtual-virtual scattering ($m=0,1,2,3$, $A,B=u,d,s,c$)
\bea
\sigma_{\rm vac}^{\gamma^*\gamma^*}(x,Q^2,P^2)=
{(4\pi^2 \alpha_{em})^{2} \over Q^{2}P^{2}}\sum_{m} C_m\sum_{A,B}
f^{A}_m(Q^2)f^{B}_m(P^2)
\left({3x_0\over 2x_{AB}}\right)^{\Delta_m}
\nonumber\\
+\sigma^{\gamma^*\gamma^* }_{\rm soft}(x,Q^2,P^2).
\label{eq:2.51}
\eea
To make explicite the scale dependence 
 discussed in Introduction  we provide $\mu$ and $x$ defined by the  eq.(\ref{eq:1.2}) 
with two indices, $A$   and $B$,  
pointing out the flavor of both the beam and target dipoles: 
$\mu^2_{AB}=m^2_{\rho}$  for $A,B=u,d,s$ while $\mu^2_{AB}=4m^2_{c}$ if either
$A=c$ or $B=c$ and the second dipole is made of light quarks
 and  $\mu^2_{AB}=8m^2_{c}$ if $A=B=c$.

For the  DIS off  real (quasireal) photons, $P^{2}\simeq 0$, we have
 $(A=u,d,s,c)$
\beq
F_{2\gamma}(x,Q^2)=\sum_m A^{\gamma}_m\sum_{A}f^A_m(Q^2)
\left({3\over 2}{x_0\over {x_A}}\right)^{\Delta_m}+F_{2\gamma}^{\rm soft}(x,Q^2)
+F_{2\gamma}^{\rm nvac}(x,Q^2)\,,
\label{eq:2.61}
\eeq
where 
$$x_A={Q^2+\mu_A^2\over W^2+Q^2}$$
  and $\mu^2_{A}=m^2_{\rho}$ for $A=u,d,s$
 while  $\mu^2_{A}=4m^2_{c}$ for $A=c$.
 The $c\bar c$  component of the target photon wave function is strongly suppressed 
at $P^2\simeq 0$ and for all the practical purposes can be neglected
 as well as $c\bar c$ content of the target proton.
 This observation simplifies the factorization relation (\ref{eq:2.61})
 for the real photon structure function. 
 In eq.(\ref{eq:2.61}) the non-vacuum component 
denoted by $F_{2\gamma}^{\rm nvac}$ is
\beq
F_{2\gamma}^{\rm nvac}(x,Q^2)=F_{2\gamma}^{\rm had}(x,Q^2)+F_{2\gamma}^{\rm pl}(x,Q^2)
\label{eq:NONVAC}
\eeq
and the cross sections
\beq
\sigma_m^{\gamma^*}(Q^2)=
 \langle {\gamma^*_T}  | \sigma_m(r)|{\gamma^*_T}
\rangle +  \langle {\gamma^*_L}  | \sigma_m(r)|{\gamma^*_L}
\rangle\,.
 \label{eq:2.8}
\eeq
are calculated with the well known color dipole distributions in the 
transverse (T) and  longitudinal (L) photon of virtuality $Q^{2}$ derived 
in \cite{NZ91}, and the eigen SFs are defined as
usual: 
\beq
f_m(Q^2)= {Q^2\over
 {4\pi^2\alpha_{\rm em}}}
 \sigma_m^{\gamma^*}(Q^2)\, .
\label{eq:2.9}
\eeq
The factor ${3\over 2}$ in the Regge parameter derives from the point that
in a scattering of color dipole on the photon the effective dipole-dipole 
collision energy is ${3\over 2}$ of that in the reference scattering of 
color dipole on the three-quark nucleon at the same total c.m.s. energy $W$. 
The analytical formulas for the  eigen-SFs $f_m(Q^2)$  and $f^c_m(Q^2)$ are 
found in the Appendix. Here as well as in all our previous calculations we
 put $m_c=1.5$ GeV.
 We do not need any new parameters compared to 
those used
in the description of DIS and real photoabsorption on protons 
\cite{JETPLett,DER,PION} (for an alternative approach see \cite{ASK,NAVEL}),
 the results
for the expansion parameters $A_{m}^{\gamma}=C_m\sigma_m^{\gamma}$, 
$C_m=1/\sigma_m^p$ 
and $\sigma_{m}^{\gamma}\equiv\sigma_{m}^{\gamma^*}(0)$
are summarized in the Table~1.

We recall that because of the diffusion in color dipole space,
 exchange by perturbative gluons contributes also to interaction
 of large dipoles
$r>R_c$ \cite{NZZJETP}. However at moderately large Regge parameter this hard 
interaction driven effect is still small. For this reason in what follows we
 refer to terms $m=0,1,2,3$ as hard contribution
as opposed to the genuine soft interaction.

\vspace{0.25cm}
\begin{center}
Table 1. CD BFKL-Regge expansion parameters.\\
\begin{tabular}{|c|c|c|c|c|c|c|} \hline
$m$& $ \Delta_m$ & $\sigma_m^{p},{\rm\, mb}$ &
 $C_m  ,{\rm\, mb^{-1}}$& $A_m^{\gamma}/\alpha_{\rm em}$ &
 $\sigma_m^{\gamma},{\rm\, \mu b}$ &$\sigma_m^{\gamma\gamma},{\rm\, nb}$ \\ \cline{1-7}
0 &0.402  & 1.243  &0.804 & 0.746& 6.767 &    36.84\\ \cline{1-7}
1 &0.220  &0.462  &2.166 &  0.559 & 1.885& 7.69   \\ \cline{1-7}
2 &0.148  &0.374  &2.674 &  0.484 & 1.320& 4.65   \\ \cline{1-7}
3 &0.06   &3.028  &0.330 &  0.428 & 9.456& 29.53   \\ \cline{1-7}
{\rm soft} & 0.&31.19 &0.0321& 0.351  &79.81& 204.2 \\ \cline{1-7}
\end{tabular}
\end{center}

\section{Isolating the soft plus rightmost hard BFKL pole in highly virtual-virtual 
$\gamma^{*}\gamma^{*}$ scattering}

We start with the theoretically cleanest case of the highly virtual
photons, $P^{2},Q^{2}\gg $1 GeV$^2$ and focus on the vacuum exchange component
of the total cross section. 
 The CD BFKL approach with asymptotic freedom predicts 
uniquely that subleading eigen SFs have a node
at $Q^{2}\sim 20$ GeV$^2$ in which region of $Q^{2}$ the rightmost
hard pole contribution will dominate. This suppression of the
 subleading hard background is shown in Fig.~1, in which 
we plot the ratio ($m=0,1,2,3,soft$)
$$
r_{m}(Q^{2}) ={\sigma_m^{\gamma^*\gamma^*}({3\over 2}x_0,Q^2,Q^2)\over 
\sigma_{\rm vac}^{\gamma^*\gamma^*}({3\over 2}x_0,Q^2,Q^2)}\, ,
$$ 
which defines the relative size of different contributions to 
$\sigma_{\rm vac}^{\gamma^*\gamma^*}$  at $x={3\over 2}x_{0}$. At this value of $x$
the contribution of subleading hard BFKL poles remains marginal in a 
broad range of $Q^{2}$, although the contribution from the single-node 
component $m=1$ becomes substantial at $Q^{2} \gsim 10^{3}$ GeV$^2$.

The soft-pomeron exchange contributes substantially over all $Q^2$
and dominates at $Q^{2} \lsim 1$ GeV$^2$. However, at very large $W\sim 100$ GeV 
of the practical interest at LEP and LHC, such small values of $Q^{2}$ 
correspond to very small $x$, where the
soft and subleading hard contributions are Regge suppressed by the 
factor $\left({x/x_{0}}\right)^{\Delta_{\Pom}}$ and 
$ \left({x/ x_{0}}\right)^{0.5\Delta_{\Pom}}$, respectively.
The latter is clearly seen from Fig.~2
where the effective pomeron intercept
\beq
\Delta_{eff}=-{\partial\log{\sigma_{\rm vac}^{\gamma^*\gamma^*}}\over{\partial\log{x}}}
\label{eq:DELEFF}
\eeq
is presented  for the diagonal case $Q^2=P^2$ at three different values of $W$.

According to the results shown in Fig.~1 the dominance of the soft plus  rightmost 
hard BFKL pomeron exchange in  virtual-virtual $\gamma^{*}\gamma^{*}$ 
scattering holds in a very broad range of $Q^{2},P^{2} \lsim 500$
GeV$^{2}$  which nearly exhausts the interesting kinematical region
at LEP200 and NLC. 
The quality of the leading hard pole  plus soft approximation (LHSA)
can be judged also from Fig.~3 for the diagonal case of $Q^{2}=P^{2}$, 
in which we show separately the soft component of the cross section
(the dashed curve). The point that the contribution from subleading 
hard BFKL exchange is marginal is clear from the finding that approximation 
of soft-pomeron plus the rightmost hard BFKL exchange (LHSA) shown
by long-dashed curve nearly exhausts the result from the complete CD
BFKL-Regge expansion for vacuum exchange.

Recently the L3 collaboration \cite{L3SIGY} reported the first experimental 
evaluation of the vacuum exchange in equal virtuality $\gamma^{*}
\gamma^{*}$ scattering. Their procedure of subtraction of the 
non-vacuum reggeon and/or the Quark Parton Model contribution
is  described  in \cite{L3SIGY}, arguably 
the subtraction
uncertainties are marginal within the present error bars. In Fig.~4 we
compare our predictions to the L3 data. The experimental data and
theoretical curves are shown vs. the variable $Y=\log(W^2/\sqrt{Q^2P^2})$. 
The  virtuality of two photons varies in the range of $1.2\,GeV^2<Q^2,P^2
<9\,GeV^2$ ($\langle  Q^2,P^2\rangle=3.5\,$GeV$^2$) at $\sqrt{s}\simeq 91\,GeV$
and $2.5\,GeV^2<Q^2,P^2<35\,GeV^2$ at $\sqrt{s}\simeq 183\,GeV$
($\langle  Q^2,P^2\rangle=14\,$GeV$^2$).  We applied to the theoretical
cross sections the same averaging procedure as described in \cite{L3SIGY}.
The solid curve is a result of the complete BFKL-Regge expansion for the
vacuum exchange, the long-dashed curve is a sum of the rightmost hard BFKL
exchange and soft-pomeron exchange. Shown by the dashed line is the soft pomeron
contribution. The agreement of our estimates 
with the experiment is good, the contribution of subleading hard BFKL 
exchange is negligible within the experimental error bars.

In Fig.~5 we compare our predictions for the vacuum exchange contribution to
$\sigma^{\gamma^*\gamma^*}(Y)$ with  recent OPAL Collaboration 
measurements \cite{OPALY}. In the applicability region of our approach 
corresponding to $Y\gsim Y_0=\log(2/3x_0)\simeq 3$ the agreement with data
is good. The discrepancy at smaller $Y$ may indicate  
significant 
non-vacuum contributions vanishing at large $Y$.

The early calculations \cite{BRODSOP,BART1,BART2} of the perturbative 
vacuum component of
$\sigma ^{\gamma^*\gamma^*}$ used the approximation $\alpha_{S}=const$
which predicts the $P^{2},Q^{2}$-dependence different from our result
 for CD BFKL approach with running $\alpha_{S}$. Detailed
comparison with numerical results by Brodsky, Hautmann and  Soper (BHS)
 \cite{BRODSOP}
is reported by the L3 Collaboration \cite{L3SIGY}, which founds that 
BHS formulas overpredict $\sigma_{\rm vac}^{\gamma^*\gamma^*}$ substantially.
In \cite{BART2}  the same perturbative
 fixed-$\alpha_S$ BFKL model  
 with massive $c$-quark
has been considered. At  $\langle Q^2\rangle=14$ GeV$^2$ and moderately small $x$,
$x\gsim 3.10^{-2}$, the model
is in agreement with the L3 data but at smaller $x$, already at $x\sim 7.10^{-3}$,
it substanially overpredicts $\sigma_{\rm vac}^{\gamma^*\gamma^*}$.  At
 $\langle Q^2\rangle=3.5$ GeV$^2$ the results \cite{BART2}  are substantially
 above the  L3 data over all $x$.


\section{Virtual-real $\gamma^{*}\gamma$ scattering: the rightmost 
hard BFKL pole in the photon SF}

The discussion of the photon SF  follows closely that
of the proton and pion SF's in \cite{JETPLett,DER,PION}. Our normalization of
eigen-functions is such that the vacuum (sea) contribution to the proton 
SF ($m={\rm soft},0,1,..,3$)
\beq
F_{2p}(x,Q^2)=\sum_m 
f_m(Q^2)
\left({x_0\over {x}}\right)^{\Delta_m}
\label{eq:4.1}
\eeq
has the CD BFKL-Regge expansion coefficients $A_{m}^{p}=      1$.
There is a fundamental point that the distribution of small-size color 
dipoles in the photon is enhanced compared to that in the proton 
\cite{PION} which enhances the importance of the rightmost hard 
BFKL exchange. Indeed, closer inspection of expansion coefficients 
$A_{m}^{\gamma}$ shown in table~1 reveals that subleading hard BFKL
exchanges are suppressed with respect to the leading one
 by the factor $\simeq$ 1.5, whereas the
soft-pomeron exchange contribution is suppressed by the factor $\simeq 2$.

Our predictions for the photon SF are parameter-free 
and are presented in 
Fig.~6. At moderately small $x\sim 0.1$ there is a substantial non-vacuum reggeon 
exchange contribution 
from DIS off hadronic ($q\bar q$) component of the target photon wave function
   which
can be regarded as well constrained by the large $x$ data. We use
here the parameterizations presented above (eqs.(\ref{eq:HAD},\ref{eq:PL},
\ref{eq:NONVAC})).
 The solid curve shows the result from
the complete BFKL-Regge expansion the soft-pomeron (the dashed curve)
and quasi-valence (the dot-dashed curve) components included, the dotted 
curve shows the rightmost hard BFKL (LH) plus soft-pomeron (S) plus 
non-vacuum  (NV) approximation (LHSNVA). A comparison of the solid and 
dotted curves shows clearly that subleading hard BFKL exchanges are
numerically small in the experimentally interesting region of $Q^{2}$,
the rightmost hard BFKL pole exhausts the hard vacuum contribution 
for $2 \lsim  Q^{2} \lsim 100\, $ GeV$^{2}$. The nodal properties of
subleading hard BFKL SFs are clearly seen: LHSNVA
underestimates $F_{2\gamma}$ slightly at $Q^{2}\lsim 10$ GeV$^2$ and
overestimates $F_{2\gamma}$ at $Q^{2}\gsim 50$ GeV$^2$. For still
another illustration of the same nodal property of subleading hard
components see Fig.~7 in which we show the vacuum component of
virtual-real total cross section $\sigma_{tot}^{\gamma^{*}\gamma}$
as a function of $Q^{2}$ at fixed $W$. As seen from Fig.~1, the
soft contribution rises towards small $Q^{2}$, but this rise is
compensated to a large extent by the small-$x$ enhancement of the 
rightmost hard BFKL contribution by the large Regge factor 
$\left({x_0\over x}\right)^{\Delta_{\Pom}}$. For this region
the soft background (the dashed curve) remains marginal over the 
whole range of $Q^{2}$. Because of the node effect, the $m=1$
subleading component changes the sign and becomes quite substantial 
at very large $Q^{2}$ and moderately small $x$.

Recently the  L3 and OPAL  collaborations reported the first experimental
data on the photon SF at sufficiently small-x \cite{L3,OPAL}.
These data are shown in Fig.~6 and are in good agreement with the
predictions from the CD BFKL-Regge expansion. A comparison with the 
long-dashed curve which is the sum of the rightmost hard BFKL and soft
exchanges shows that the experimental data are in the region of 
$x$ and $Q^{2}$  still affected by non-vacuum reggeon (quasi-valence) exchange
(dot-dashed line),
going to smaller $x$ and larger $Q^{2}$ would improve the sensitivity
to pure vacuum exchange greatly. 

 In order to give a crude idea on finite-energy effects at large $x$
and not so large values of the Regge parameter we stretch the theoretical
curves a bit to $x\gsim x_{0}$  
multiplying the BFKL-Regge expansion result by the 
purely phenomenological factor $(1-x)$ motivated by the familiar 
behavior of the gluon SF of the photon  
$\sim (1-x)^{n}$ with the exponent $n\sim 1$.


\section
{The real-real $\gamma\gamma$ scattering}

We recall that because of the well known BFKL diffusion in color 
dipole space,  exchange by perturbative gluons contributes also to 
interaction of large dipoles $r>R_c$ \cite{NZZJETP}. As discussed
in \cite{PION} this gives rise to a substantial rising component of
hadronic and real photoabsorption cross sections and a scenario 
in which the observed rise of hadronic and real photon cross sections 
is entirely due to this intrusion of hard scattering. This is a 
motivation behind our choice of intercept $\Delta_{\rm soft}=0$ for
soft pomeron exchange. Furthermore, in order to make this picture 
quantitative one needs to invoke strong absorption/unitarization 
to tame too  a  rapid growth of large dipole component of hard BFKL
the dipole cross section. The case of real-real $\gamma\gamma$ 
scattering is not an exception and the above discussed
enhancement of small dipole configurations in photons compared to
hadrons predicts uniquely that the hard BFKL exchange component of 
real-real $\gamma\gamma$ scattering will be enhanced compared to 
proton-proton and/or pion-proton scattering. This is clearly seen
from table~1 in which we show the coefficients 
\beq
\sigma_{m}^{\gamma\gamma} = \sigma_{m}^{\gamma}\sigma_{m}^{\gamma}C_{m}
\label{eq:5.1}
\eeq
of the expansion for the vacuum exchange component of the total
$\gamma\gamma$ cross section ($m=0,1,2,3,soft$) 
\beq
\sigma_{\rm vac}^{\gamma\gamma}=\sum_{m}\sigma_{m}^{\gamma\gamma}
\left({W^{2} x_{0}\over m_{\rho}^{2}}\right)^{\Delta_{m}}\, .
\label{eq:5.2}
\eeq
One has to look at the soft-hard hierarchy of 
$\sigma_{m}^{\gamma\gamma}$ and $\sigma_{m}^{\gamma},\sigma_{m}^{p}$
in the counterparts of (\ref{eq:5.2}) for $\gamma p$ and $pp$ scattering.
This enhancement of hard BFKL exchange is confirmed by 
simplified vacuum pole plus non-vacuum reggeon exchange fits to
real-real $\gamma\gamma$ total cross section: the found intercept
of the effective vacuum pole $\epsilon^{\gamma\gamma} \approx 0.21$ is 
much larger than $\epsilon \approx 0.095$ from similar fits to the 
hadronic cross section data. In Fig.~8 we compare our predictions 
from the CD BFKL-Regge factorization for the single-vacuum exchange 
contribution to real-real $\gamma\gamma$ scattering with the 
recent experimental data from the OPAL collaboration 
\cite{OPALGG} and \cite{PDG}. The theoretical curves 
are in the right ballpark, but the truly quantitative discussion
of total cross sections of soft processes requires better
understanding of absorption/unitarization effects.


\section{Regge factorization in $\gamma^*\gamma^*$ and $\gamma\gamma$
scattering} 

If the vacuum exchange were an isolated Regge pole, the well known 
Regge factorization would hold for asymptotic cross sections
\cite{Gribov}
\beq
\sigma_{tot}^{bb}\sigma_{tot}^{aa}=
\sigma_{tot}^{ab}\sigma_{tot}^{ab}\, .
\label{eq:6.1}
\eeq
In the CD BFKL approach such a Regge factorization holds for each
term in the BFKL-Regge expansion for vacuum exchange, but evidently 
the sum of 
factorized terms does not satisfy the factorization (\ref{eq:6.1}).
One can hope for an approximate factorization only provided one
single term to dominate in the BFKL-Regge expansion. Though 
corrections to the exact factorization still exist even for the single pole
exchange because of the light $q\bar q$ and charm $c\bar c$ mass scale difference
discussed above.  

One such case is real-real $\gamma\gamma$ scattering dominated
by soft-pomeron exchange (the factorization of  the soft  on-shell  amplitudes
though never proved gained strong support from the  high-energy Regge-phenomenology).
 For this reason the CD BFKL-Regge 
expansion which reproduces well the vacuum exchange components of the
$pp$ and $\gamma p$ scattering can not fail for the vacuum component
in real-real $\gamma\gamma$ scattering. The rise of the contribution
of hard-BFKL exchange breaks the Regge factorization
relation 
\beq
R_{\gamma\gamma}= {\sigma_{\rm vac}^{\gamma\gamma}\sigma^{p p}\over 
\sigma_{\rm vac}^{\gamma p}\sigma_{\rm vac}^{\gamma p}} = 1\, ,
\label{eq:6.2}
\eeq
which would restore at extremely high energies such that
the rightmost hard BFKL exchange dominates. This property is
illustrated in Fig.~9 where we show our evaluation of $R$ for 
single-vacuum component of total cross sections entering 
(\ref{eq:6.1}). At moderately high energies naive factorization
breaks but the expected breaking is still weak, $\lsim 20\%$.
This curve must not be taken at face value 
for $W\gsim $0.1-1 TeV because of likely strong absorption
effects, but the trend of $R$ being larger than unity and rising with
energy should withstand unitarity effects.
The second case is highly virtual-virtual $\gamma^*\gamma^*$
scattering.
 As we emphasized in section 3, here the CD BFKL
approach predicts uniquely that because of the nodal property
of subleading eigen SFs the the superposition of soft and rightmost
hard BFKL poles dominate the vacuum exchange in a broad range of 
$Q^{2},P^{2} \lsim 10^3$ GeV$^{2}$.

 The above discussion suggests clearly
that different cross sections must be taken at the same value
of $x^{-1}=W^2/(Q^2+P^2)$,  in which case the vacuum 
components of $\gamma^*\gamma^*$ scattering at $Q^2,P^2\gg 4m_c^2$ and 
$x\ll x_0$ would satisfy
\beq
 R_{\gamma^*\gamma^*}(x)={ {[\sigma^{\gamma^*\gamma^*}(x,Q^2,P^2)]^2}\over
{\sigma^{\gamma^*\gamma^*}(x,Q^2,Q^2)\sigma^{\gamma^*\gamma^*}(x,P^2,P^2)}}
 =1.
\label{eq:6.3}
\eeq

In accordance to the results shown in Fig.~1, the soft
 exchanges break the factorization relation (\ref{eq:6.3}).
The breaking is quite substantial at moderate $x=0.01$ 
(dotted line in Fig.~10),
 and breaking effects disappear
rapidly, $\sim x^{\Delta_{0}}$, as $x\to 0$. If the vacuum 
singularity were the Regge cut as is the case in approximation
$\alpha_S=$const, then restoration of factorization is much slower,
cf. our Fig.~10 and Fig.~9 in \cite{BRODSOP}.

For an obvious reason that the soft-pomeron exchange is so predominant
in real photon scattering, whereas the soft plus rightmost hard BFKL exchange is 
outstanding in virtual-virtual and real-virtual photon-photon scattering,
it is ill advised to look at factorization ratio    
$R_{\gamma^*\gamma^*}(W)$ when one of the photons is quasireal, $P^{2}\sim 0$.
In this limit one would find strong departures of $R_{\gamma^*\gamma^*}(W)$ from unity.
 For precisely the same reason 
of predominance of soft-pomeron exchange in $pp$ scattering vs. nearly 
dominant rightmost hard BFKL pole exchange in DIS at small $x$ and 
5-10$\lsim Q^{2} \lsim $ 100 GeV$^{2}$, see \cite{PION}, the naive 
factorization estimate
\beq
\sigma^{\gamma^*\gamma^*}(W,Q^2,P^2)\approx {\sigma^{\gamma^*p}(W,Q^2)
\sigma^{\gamma^*p}(W,P^2) \over \sigma^{pp}(W)}
\label{eq:6.6}
\eeq
would not make much sense.

\section{Conclusions}

We explored the consequences for small-$x$ photon SFs 
$F_{2\gamma}(x,Q^{2})$ and high-energy two-photon cross sections 
$\sigma^{\gamma^*\gamma^*}$ and $\sigma^{\gamma\gamma}$   from the color 
dipole BFKL-Regge factorization. Because of the nodal properties of eigen
SFs of subleading hard BFKL exchanges the CD BFKL approach 
predicts uniquely that the vacuum exchange is strongly dominated by
the combination of soft plus  rightmost hard BFKL pole exchanges
 in a very broad range of photon
virtualities $Q^{2},P^{2}$ which includes much of the kinematical domain
attainable at LEP200 and NLC.
 Starting with
very restrictive perturbative two-gluon exchange as a  boundary condition 
for BFKL evolution in the color dipole basis and having fixed the
staring point of BFKL evolution in the early resulting CD BFKL-Regge 
phenomenology of the proton SF, we presented parameter-free
predictions for the vacuum exchange contribution to the photon structure
function which agree well with OPAL and L3 determinations. A good agreement
is found between our  predictions for the energy and photon 
virtuality  dependence of the photon-photon cross section 
$\sigma^{\gamma^*\gamma^*}(W,Q^2,P^2)$ and the recent data taken by the 
L3 Collaboration. We commented on the utility of Regge factorization tests
of the CD BFKL-Regge expansion.\\

{\bf Acknowledgments: } This work was partly supported by the grants
INTAS-96-597 and INTAS-97-30494 and DFG 436RUS17/11/99.

\vspace{0.5cm}

\section{Appendix}
\subsection{CD BFKL all flavor eigen-SF}
In the early discussion of DIS off protons
 the results of numerical solutions of the CD BFKL equation
 for the all flavor ($u+d+s+c$) eigen-SF
$f_m(Q^2)$ were parameterized as
\beq
f_0(Q^2)=
a_0{R_0^2Q^2\over{1+ R_0^2Q^2 }}
\left[1+c_0\log(1+r_0^2Q^2)\right]^{\gamma_0}\,,
\label{eq:F20}
\eeq

\beq
f_m(Q^2)=a_m f_0(Q^2){1+R_0^2Q^2\over{1+ R_m^2Q^2 }}
\prod ^{m}_{i=1}\left(1-{z\over z^{(i)}_m}\right)\,,\,\, m\geq 1\,,
\label{eq:FN}
\eeq
where $\gamma_0={4\over {3\Delta_0}}$ 
and
\beq
z=\left[1+c_m\log(1+r_m^2Q^2)\right]^{\gamma_m}-1 ,\,\,\,
\gamma_m=\gamma_0 \delta_m\,.
\label{eq:ZFN}
\eeq

 The parameters  tuned to reproduce
 the numerical results for $f_m(Q^2)$ at $Q^2\lsim 10^5\, GeV^2$
are listed in the Table 2.

\vspace{0.5cm}
\begin{center}
Table 2. CD BFKL-Regge the all flavor SF parameters.\vspace{0.45cm}\\

\begin{tabular}{|l|l|l|l|l|l|l|l|l|} \hline
$m$ & $a_m$  & $c_m$ & $r_m^2\,,$ ${\rm GeV^{-2}}$ &
$ R_m^2\,,$ ${\rm GeV^{-2}}$ &
$z^{(1)}_m$ & $z^{(2)}_m$ & $z^{(3)}_m$ &  $\delta_m$ \\ \cline{1-9}
0 & 0.0232  & 0.3261&1.1204&2.6018& & & & 1. \\ \cline{1-9}
1 & 0.2788 &0.1113&0.8755&3.4648&2.4773 &  &  &1.0915 \\ \cline{1-9}
2 & 0.1953 &0.0833&1.5682&3.4824 &1.7706 &12.991&  &1.2450  \\ \cline{1-9}
3 &1.4000  &0.04119 &3.9567 & 2.7706    &0.23585 &0.72853&1.13044   &0.5007 \\ \cline{1-9}
{\rm soft}&0.1077 & 0.0673& 7.0332 & 6.6447 &   &  & &      \\  \cline{1-9}
\end{tabular}
\end{center}
\vspace{0.5cm}

The soft component of the proton structure function as derived from
$\sigma_{\rm soft}(r)$ taken from \cite{JETPVM} is parameterized as follows
\beq
f_ {\rm soft}(Q^2)= 
{a_{\rm soft}R^2_{\rm soft}Q^2\over{1+R^2_{\rm soft} Q^2 }}
\left[1+c_{\rm soft}\log(1+r^2_{\rm soft}Q^2)\right]\,,
\label{eq:FSOFT}
\eeq
with parameters cited in the Table 2.

 The cross section $\sigma^{\gamma^*\gamma^*}_{\rm soft}(Q^2,P^2)$
obtained by the continuation of the above 
$$\sigma^{\gamma^*p}_{\rm soft}={4\pi^2\alpha_{em}\over Q^2}f_{\rm soft}(Q^2)$$
into $Q^2,P^2$-plane reads
\beq
\sigma^{\gamma^*\gamma^*}_ {\rm soft}(Q^2,P^2)= 
{\sigma^{\gamma\gamma}_{\rm soft}\over{1+R^2_{\rm soft} (Q^2+P^2)}}
\left[1+c_{\rm soft}\log\left(1+{r^2_{\rm soft}Q^2\over{1+r^2_{\rm soft}P^2}}
+{r^2_{\rm soft}P^2\over{1+r^2_{\rm soft}Q^2}}\right)\right]\,,
\label{eq:SGGSTAR}
\eeq
with parameters cited in the Table 2 and
the on-shell cross section 
\beq
\sigma^{\gamma\gamma}_{\rm soft}= 
\left[4\pi^2\alpha_{em}a_{\rm soft}R^2_{\rm soft}\right]^2
{1\over \sigma^{pp}_{\rm soft}}\,. 
\label{eq:SGGSOFT}
\eeq

\subsection{CD BFKL charm eigen-SF}

In practical evaluations one needs the
  charm  eigen-SF,
$f^{c}_m(Q^2)$.  For the rightmost hard BFKL pole it is of the form
\beq
f^c_0(Q^2)=
a_0{R_0^2Q^2\over{1+ R_0^2Q^2 }}
\left[1+c_0\log(1+r_0^2Q^2)\right]^{\gamma_0}\,,
\label{eq:F20C}
\eeq
where $\gamma_0=4/(3\Delta_0)$, while for the sub-leading hard BFKL poles
\beq
f^c_m(Q^2)=a_m f_0(Q^2){1+ K_m^2Q^2\over{1+ R_m^2Q^2 }}
\prod ^{m_{max}}_{i=1}\left(1-{z\over z^{(i)}_m}\right)\,,\,\, m\geq 1\,,
\label{eq:FNC}
\eeq
where $m_{max}=$min$\{m,2\}$
and
\beq
z=\left[1+c_m\log(1+r_m^2Q^2)\right]^{\gamma_m}-1 ,\,\,\,
\gamma_m=\gamma_0 \delta_m.
\label{eq:ZFNC}
\eeq
The parameters  tuned to reproduce  the numerical results for 
$f^{c}_m(Q^2)$ at $Q^2\lsim 10^5\, GeV^2$ are listed in the Table 3.

The soft component of the charm SF is parameterized as 
\beq
f^{c}_ {\rm soft}(Q^2)= 
{a_{\rm soft}R^2_{\rm soft}Q^2\over{1+R^2_{\rm soft} Q^2 }}
\left[1+c_{\rm soft}\log(1+r^2_{\rm soft}Q^2)\right]\,,
\label{eq:FCSOFT}
\eeq
with parameters cited in the Table 3.

\vspace{5mm}

\begin{center}
Table 3. CD BFKL-Regge charm structure functions parameters.
\begin{tabular}{|c|c|c|c|c|c|c|c|c|} \hline
$m$& $a_m$ &  $c_m$ & $r_m^2,$         & $ R_m^2,$       &$ K^2_m,$  &$z^{(1)}_m$ & $z^{(2)}_m$ & $\delta_m$ \\ 
   &       &        & ${\rm GeV^{-2}}$ & ${\rm GeV^{-2}}$&${\rm GeV^{-2}}$ &            &             &     \\ \cline{1-9}
0  &  0.02140        & 0.2619  & 0.3239 & 0.2846&          &        &  & 1. \\ \cline{1-9}
1  & 0.0782         & 0.03517  &0.0793  &0.2958 & 0.2846  &  0.2499&  & 1.9249\\ \cline{1-9}
2  & 0.00438        &0.03625   &0.0884  &0.2896 & 0.2846  &  0.0175 &3.447&1.7985\\ \cline{1-9}
3  &$-0.26313$      &2.1431    &$3.7424\cdot 10^{-2}$ &$8.1639\cdot 10^{-2}$ & 0.13087   &158.52   & 559.50 &0.62563  \\ \cline{1-9}
{\rm soft}& 0.01105  &0.3044   &0.09145 &0.1303 &         &         &     &        \\ \cline{1-9}
\end{tabular} 
\end{center}
\vspace{5mm}


\newpage

{\large\bf{ Figure Captions}\\ }

\begin{enumerate}

\item[{\bf Fig.1}]
The normalized ratio of soft-to-rightmost-hard and subleading hard-to-rightmost hard 
expansion coefficients ($m=1,2,3,{\rm soft}$)
$
r_{m}(Q^{2}) =\sigma_m^{\gamma^*\gamma^*}/ 
\sigma^{\gamma^*\gamma^*}_{\rm vac}
$ 
of the BFKL-Regge expansion for $\gamma^*\gamma^*$ scattering at  $x=x_{0}$.

\item[{\bf Fig.2}]
Predictions from CD BFKL-Regge expansion for the
effective intercept $\Delta_{eff}$, eq.(\ref{eq:DELEFF}), for the
diagonal case $Q^2=P^2$ and  $W=50, 100, 200$ GeV.

\item[{\bf Fig.3}]
Predictions from CD BFKL-Regge expansion for the vacuum exchange component 
of the virtual-virtial $\gamma^{*}\gamma^{*}$ cross section for the diagonal 
case of $Q^{2}=P^{2}$ and for  cms collision energy  W=50, 100 and 200 GeV 
(solid curves).
The Leading Hard BFKL exchnage   plus Soft-pomeron exchange Approximation 
(LHSA) is shown by the long dashed curve. 
The soft pomeron  component of the cross section is  shown separately
by the dashed curve.

\item[{\bf Fig.4}]
Predictions from CD BFKL-Regge expansion for the vacuum exchange component 
of the virtual-virtial $\gamma^{*}\gamma^{*}$ cross section for the diagonal 
case of $\langle Q^{2} \rangle = \langle P^{2}\rangle $  are confronted
to the experimental data by the L3 Collaboration \cite{L3SIGY}.
 The experimental 
data and theoretical curves are shown vs.
 the variable $Y=\log(W^2/\sqrt{Q^2P^2})$.
 The solid curve shows the result from
the complete BFKL-Regge expansion, the soft-pomeron (the dashed curve)
 component included. the long dashed  
curve shows the rightmost hard BFKL  (LH) plus soft-pomeron  (S)
 approximation (LHSA).   

\item[{\bf Fig.5}]
Predictions from CD BFKL-Regge expansion for the vacuum exchange component 
of the virtual-virtial $\gamma^{*}\gamma^{*}$ cross section for the diagonal 
case of $\langle Q^{2} \rangle = \langle P^{2}\rangle=17.9$ GeV$^2$
  are confronted
to the experimental data by the OPAL Collaboration \cite{OPALY}.
 The experimental 
data and theoretical curves are shown vs.
 the variable $Y=\log(W^2/\sqrt{Q^2P^2})$. 
The solid curve shows the result from
the complete BFKL-Regge expansion, the soft-pomeron (the dashed curve)
 component included. The long dashed line corresponds to 
the rightmost hard BFKL (LH) plus  soft-pomeron (S) approximation (LHSA).

\item[{\bf Fig.6}]
Predictions from CD BFKL-Regge expansion  for the photon SF.
 The solid curve shows the result from
the complete BFKL-Regge expansion the soft-pomeron (the dashed curve)
and valence (the dot-dashed curve) components included, the dotted 
curve shows the rightmost hard BFKL  (LH) plus soft-pomeron  (S) plus 
non-vacuum (NV) approximation (LHSNVA). The long dashed line corresponds to 
the LH plus S approximation (LHSA). Data points are from \cite{L3,OPAL} 

\item[{\bf Fig.7}]
Predictions from CD BFKL-Regge expansion for the vacuum exchange component 
of the the virtual-real $\gamma^*\gamma$ total cross section
and for  cms collision energy  W=50, 100 and 200 GeV (solid curves).
The result from the rightmost Hard BFKL (LH) plus Soft-pomeron (S)  
Approximation (LHSA) is shown by the long dashed curve. 
The soft-pomeron exchange  component of the cross section is  shown separately
by the dashed curve.

\item[{\bf Fig.8}]
Our predictions 
from the CD BFKL-Regge factorization for the single-vacuum exchange 
contribution to real-real $\gamma\gamma$ scattering are compared with the 
recent experimental data from the OPAL collaboration 
\cite{OPALGG} and \cite{PDG}.

\item[{\bf Fig.9}]
Our evaluation of 
$R_{\gamma\gamma}$
 for 
single-vacuum component of total cross sections.

\item[{\bf Fig.10}]
The factorization cross section ratio $R_{\gamma^*\gamma^*}(x)$ at fixed $x$ and
 $QP$
 as a function of $Q/P$ for $x=10^{-2}$ (dotted line),
$x=10^{-3}$  (long-dashed) and $x=10^{-4}$ (dashed).
\end{enumerate}

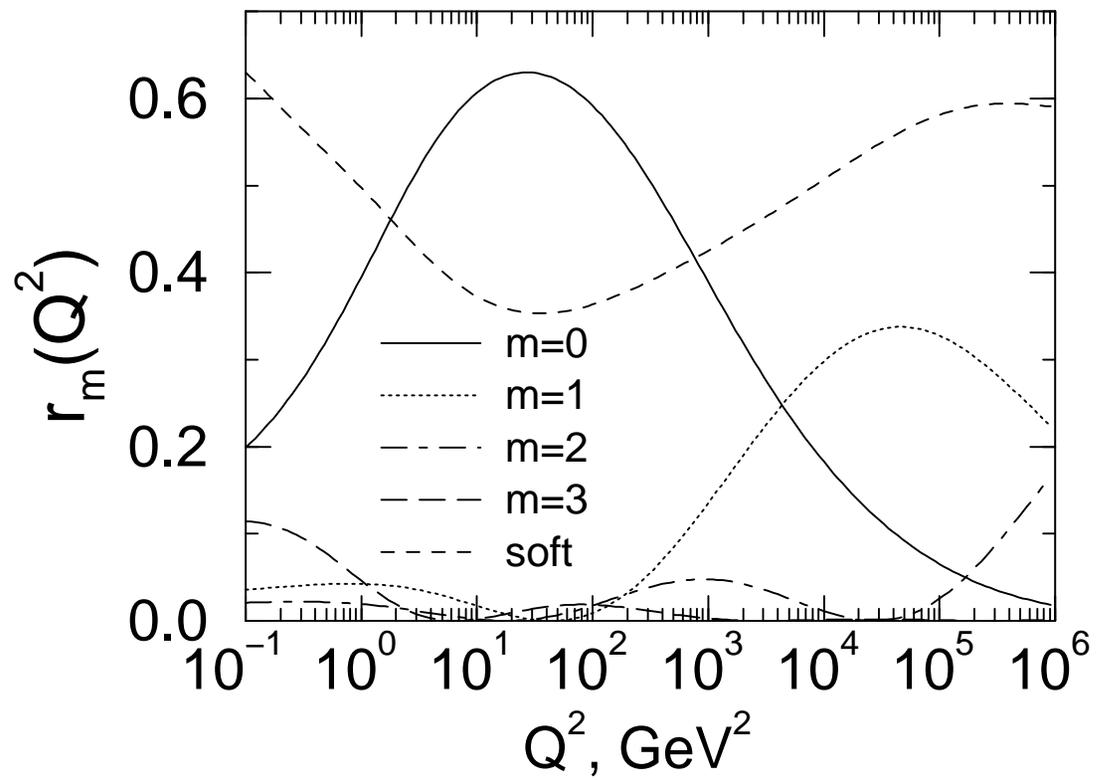
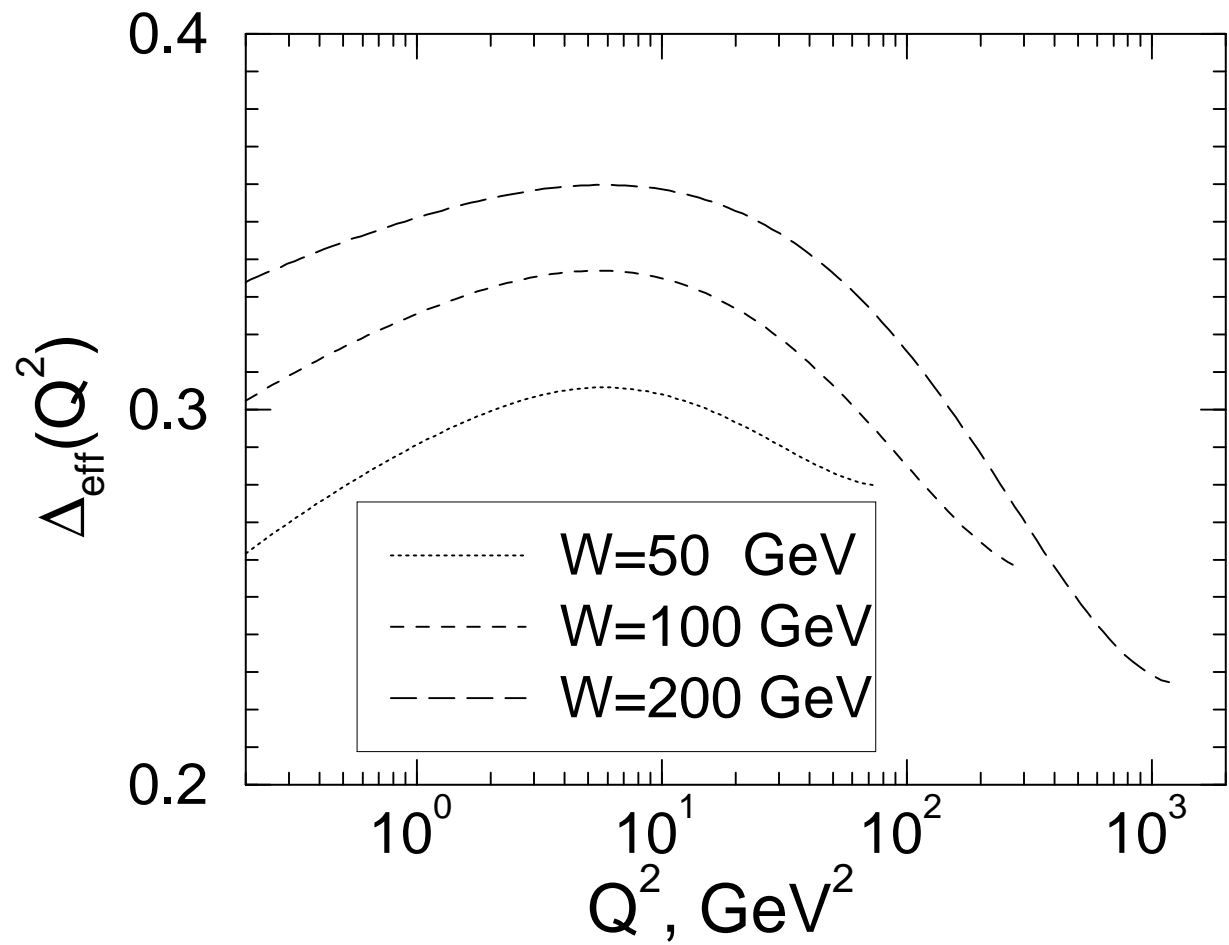
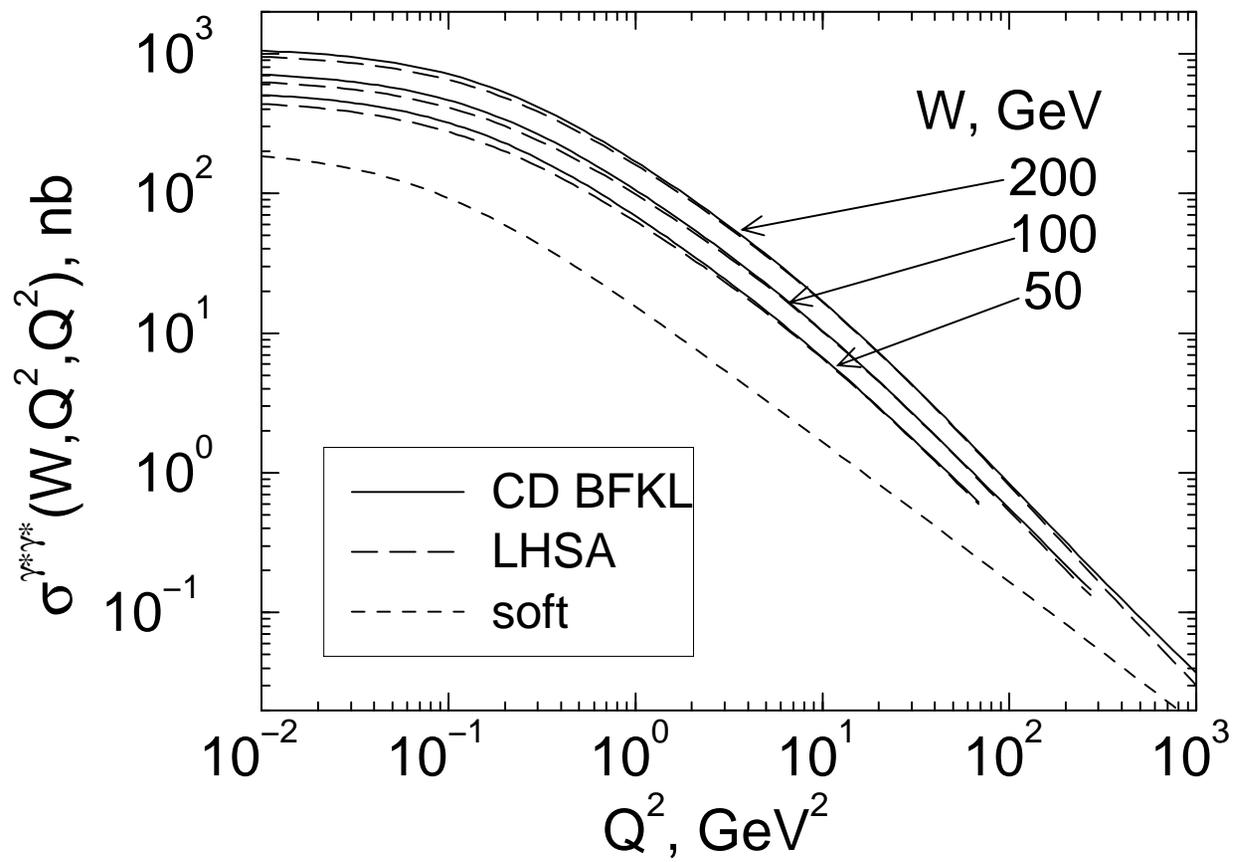
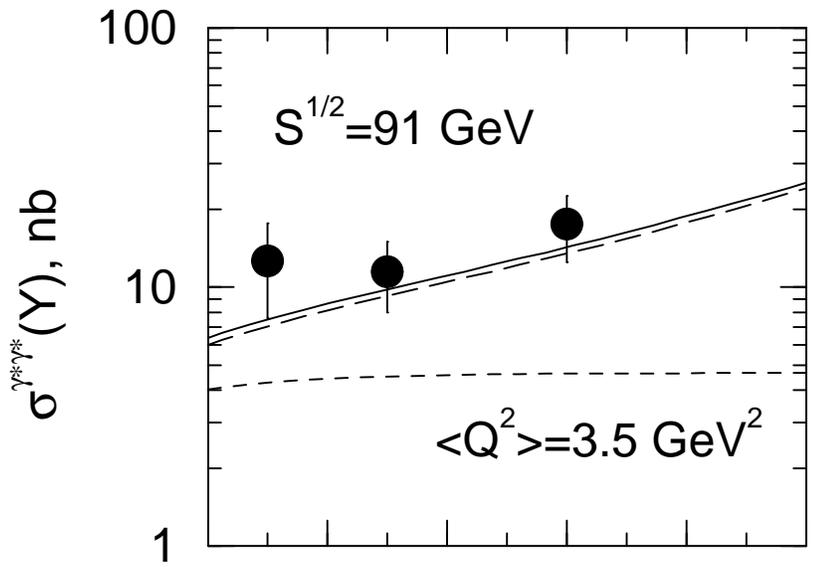
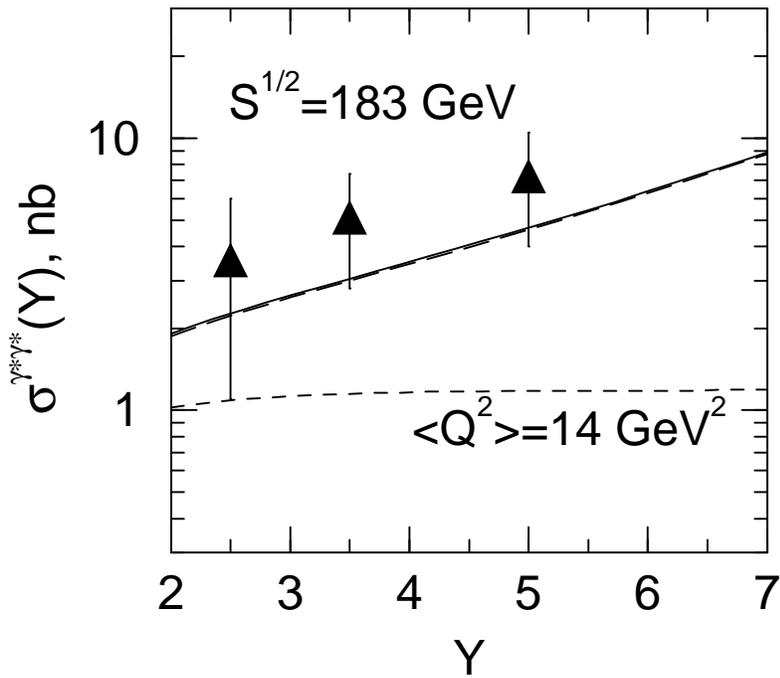
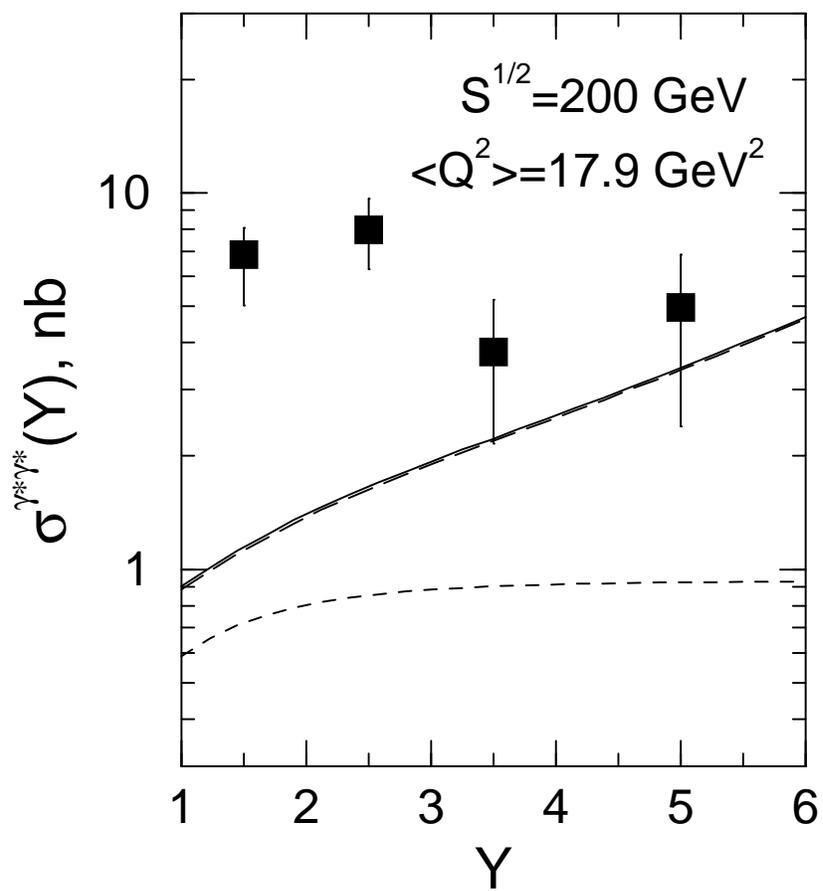
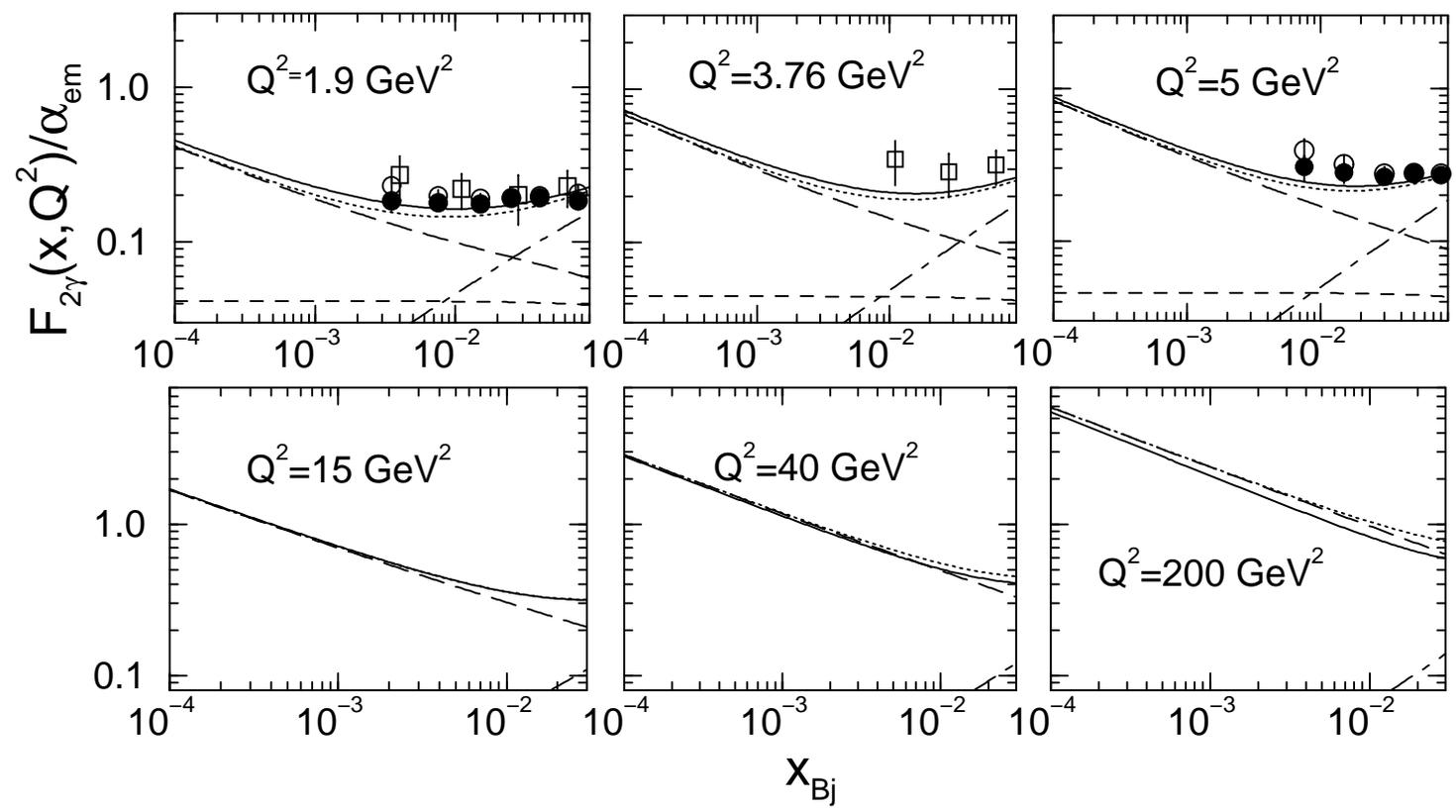
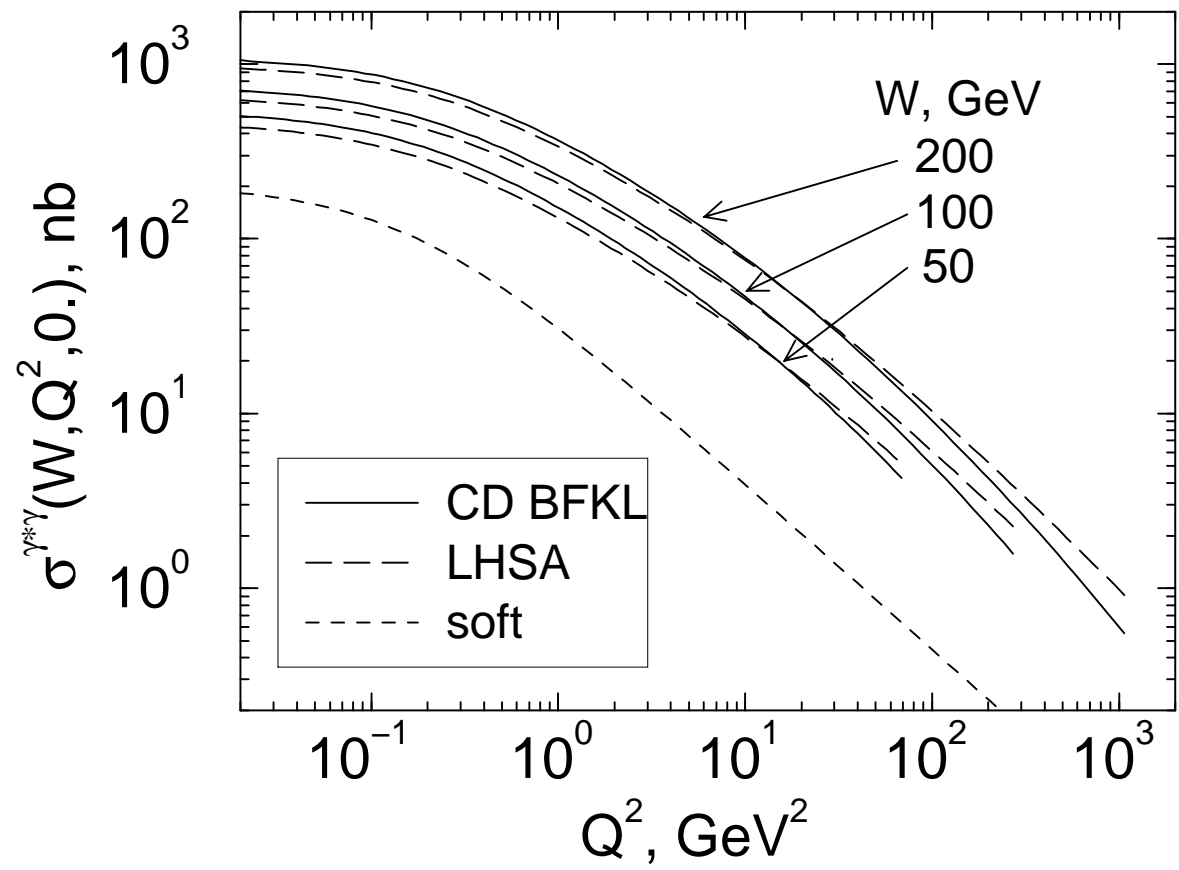
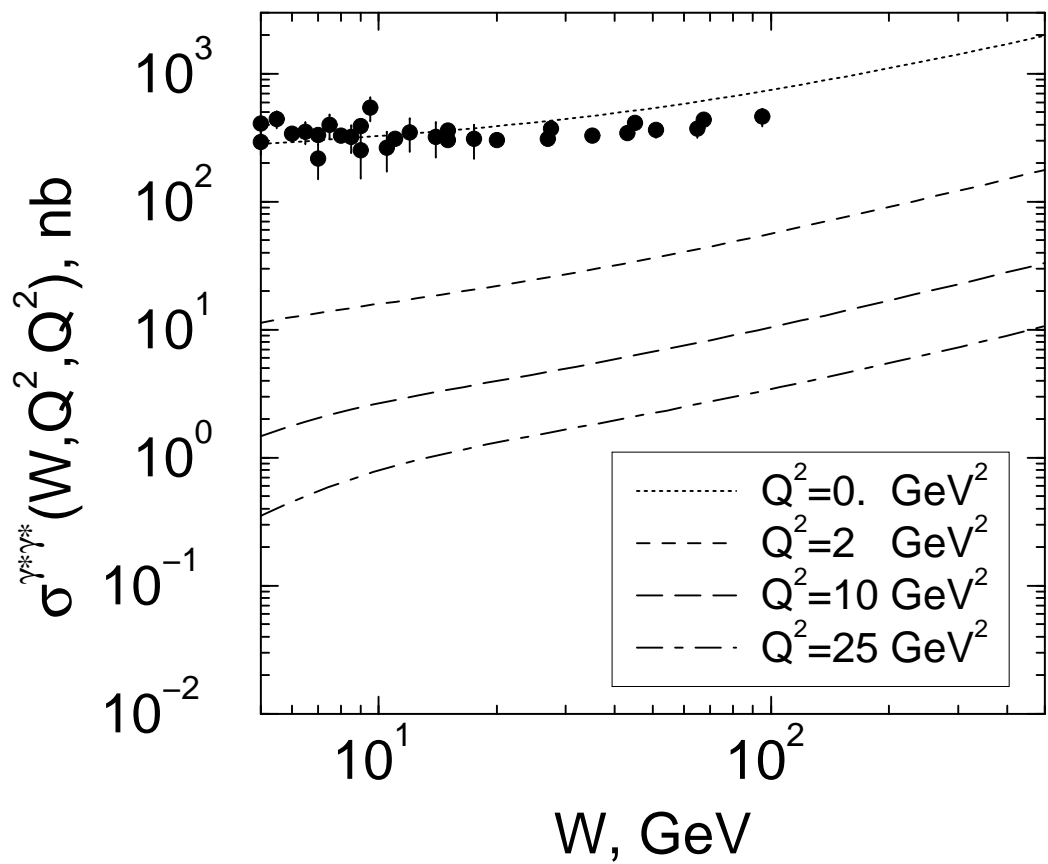
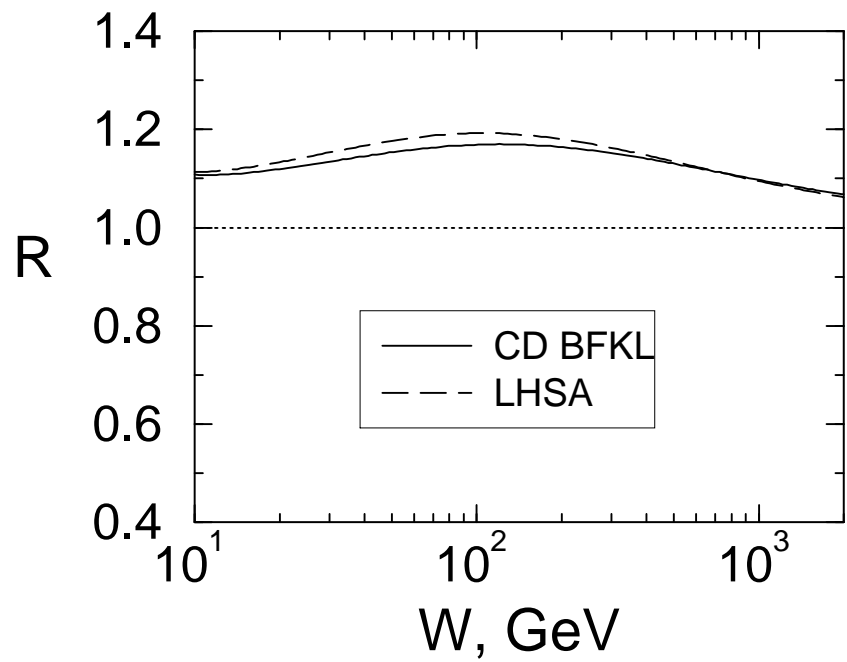
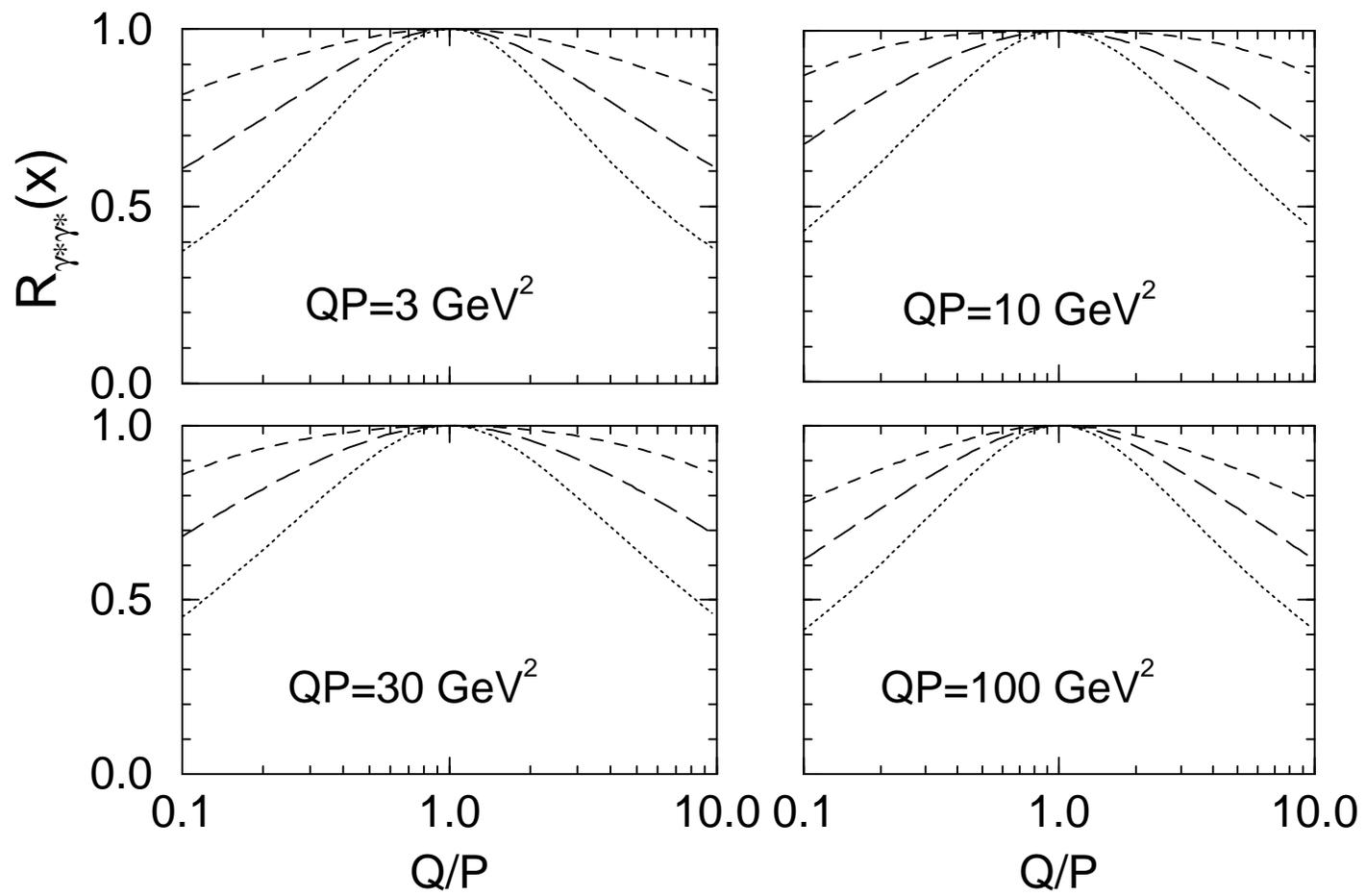
\end{document}